\documentstyle[12pt]{article}
\begin{document}
\centerline
{{\large {\bf Time-Dependent Symmetries of Variable-Coefficient}}} 
\centerline{{\large {\bf Evolution Equations and Graded Lie Algebras}}}
\vskip 2mm

\centerline {W. X. Ma\footnote{Current address: Dept. of Math., City
University of Hong Kong, Kowloon, Hong Kong; Email: 
mawx@math.cityu.edu.hk}, 
R. K. Bullough\footnote{Email: Robin.Bullough@umist.ac.uk},
P. J. Caudrey\footnote{Email: Philip.Caudrey@umist.ac.uk}  
and W. I. Fushchych\footnote{Email: math@nonlinear.kiev.ua}
}  
\centerline{Department of Mathematics, 
University of Manchester} 
\centerline{Institute of Science and Technology, 
Manchester M60
1QD, UK}   

\newtheorem{thm}{Theorem}
\newtheorem{lemma}{Lemma}
\newcommand{\R}{\mbox{\rm I \hspace{-0.9em} R}}        
\newcommand{\Z}{\mbox{\rm Z}}        
\def\be{\begin{equation}}
\def\ee{\end{equation}}
\def\ba{\begin{array}}
\def\ea{\end{array}}
\def\bea{\begin{eqnarray}}
\def\eea{\end{eqnarray}}
\def\la {\lambda}
\def \part {\partial}
\def \al {\alpha}
\def \de {\delta}

\setlength{\baselineskip}{18pt}

\begin{abstract}
Polynomial-in-time dependent symmetries are analysed for 
polynomial-in-time dependent evolution equations. 
Graded Lie algebras, especially Virasoro algebras, are used to construct
nonlinear variable-coefficient evolution equations, 
both in $1+1$ dimensions and in $2+1$ dimensions, 
which possess higher-degree polynomial-in-time 
dependent symmetries. The theory 
also provides a kind of new realisation of graded Lie algebras. Some 
illustrative examples are given.
\end{abstract}

It is well known that the usual family of KdV equations has  
polynomial-in-time dependent symmetries (ptd-symmetries)
which are only of the first-degree. 
This is because only master symmetries 
of first degree are so far found.
Moreover there are usually 
\footnote{ 
The Benjamin-Ono equation is a counter-example.} 
no higher-degree ptd-symmetries 
for time-independent integrable equations in $1+1$ dimensions;
but this may not be so in $2+1$ dimensions. 

However a form of special graded Lie algebras, namely
centreless Virasoro symmetry algebras is apparently common 
to all time-independent 
integrable equations in whatever dimensions both in the continuous case and
in the discrete case. This feature would therefore seem to be  
an important one in the discussion 
of integrability and integrable nonlinear equations. 
For the higher dimensional integrable equations, 
there may also exist still more general graded symmetry Lie algebras. 

The purpose of the present paper 
is to discuss ptd-symmetries for evolution equations
with polynomial-in-time dependent 
coefficients (conveniently expressed in terms of monomials in $t$ as in eqn. 
(\ref{pee}) below).
We provide a purely algebraic structure for constructing such 
integrable equations with these forms of symmetries.
This way we show there do exist
integrable equations in $1+1$ dimensions  
which possess these forms of symmetries and we construct actual examples.
Graded Lie algebras, and especially centreless Virasoro algebras, are used 
for these constructions. In consequence new features are
extracted from the graded Lie algebras which provide
new realisations of  
these algebras and most particularly of the centreless Virasoro algebras.

We first define a symmetry for an evolution equation, linear and nonlinear
\cite{Olver} \cite{Fuchssteiner1983} \cite{ChenL} \cite{MaF} \cite{ChengLB}. 
For a given evolution equation 
$u_t=K(u)$, a vector field $\sigma (u)$ is called its symmetry if
$\sigma (u)$ satisfies its linearized equation
\be \frac {d \sigma (u)}{dt}=K'[\sigma ],\ {\rm i.e.} \ \frac {\part \sigma}
{\part t}=[K,\sigma ],\label{sd}\ee
where the prime and $[\cdot ,\cdot]$ denote the Gateaux derivative and 
the Lie product
\be K'[S]=\frac{\part }{\part \varepsilon }K(u+\varepsilon S)
|_{\varepsilon =0},\ [K,\sigma ]=K'[\sigma ]-\sigma '[K],\ee 
respectively.
Of course, a symmetry $\sigma$ 
may also depend explicitly on the time variable $t$. 
For example,  $\sigma $ may be of 
  polynomial type in $t$, i.e.  
\be \sigma (t,u)=\sum_{j=0}^{n}   
\frac{t^j}{j!}S_j(u)=S_0+tS_1+\cdots +\frac {t^n}{n!}S_n,
\label{ps} \ee
where the vector fields $S_j(u),\ 0\le j\le n,$ do not depend explicitly
on the time variable $t$. 

If we consider a variable-coefficient 
evolution equation $u_t=K(t,u)$ of the form
\be u_t=K(t,u)=\sum_{i=0}^m \frac {t^i}{i!} T_i(u)=T_0+tT_1+\cdots + 
\frac{t^m}{m!} T_m,\label{pee}\ee
where the vector fields 
$T_i(u),\ 0\le i\le m,$ do not depend explicitly on the time variable $t$, 
either, then a precise result may be obtained which states  
(\ref{ps}) is a symmetry of (\ref{pee}).
At this stage, we can have
\bea 
\frac {\part \sigma }{\part t} & =& \sum_{i=0}^n\frac{t^{i-1}}{(i-1)!}
S_i(u)=\sum_{k=0}^{n-1}\frac{t^k}{k!}S_{k+1}(u),\nonumber \\
\left [K,\sigma \right] &= & \Bigl [\sum_{i=0}^m  \frac{t^i}{i!}T_i(u),
\sum_{j=0}^n  \frac{t^j}{j!}S_j(u)\Bigr ]\nonumber \\
&=&  \sum_{k=0}^{m+n}  \frac{t^k}{k!} 
\sum_{\ba {c} \vspace{-6mm}\\{_{i+j=k}} \vspace{-2mm} \\ {_{0\le i\le m}}
\vspace{-2mm} \\ {_{0\le j\le n}} \ea } 
{k \choose i}
[T_i,S_{j}] .\nonumber
\eea 
Therefore a simple comparison of each power of $t$ in (\ref{sd}) leads to 
\bea 
&&S_{k+1}=
\sum_{\ba {c} \vspace{-6mm}\\
{_{i+j=k}} \vspace{-2mm} \\ {_{ 0\le i\le m}}
\vspace{-2mm} \\ {_{0\le j\le n}} \ea } 
{k \choose i} [T_i,S_{j}], \ 0\le k\le n-1, \label{grule1}\\
&& 
\sum_{\ba {c} \vspace{-6mm}\\
{_{ i+j=k}} \vspace{-2mm} \\ {_{ 0\le i\le m}}
\vspace{-2mm} \\ {_{0\le j\le n}} \ea } 
{k \choose i} [T_i,S_{j}]=0, \ n\le k\le m+n.   
\label{grule2}\eea 
These equalities in (\ref{grule1}) and (\ref{grule2}) 
constitute a necessary and sufficient condition to state that (\ref{ps})
is a symmetry of (\ref{pee}). If we look at them a little more, 
it may be seen that 
\bea 
&&S_1=[T_0,S_0],\nonumber \\ 
&&S_2=[T_0,S_1]+[T_1,S_0],\nonumber \\       
&& \cdots\cdots \nonumber \\
&&S_n={n-1 \choose 0}[T_0,S_{n-1}]+{n-1 \choose 1}[T_1,S_{n-2}]+\cdots +
{n-1 \choose n-1}[T_{n-1}, S_0],\nonumber        
\eea 
where $T_i=0,\ i\ge m+1$, and so
a higher-degree ptd-symmetry
$\sigma (t,u)$ defined by (\ref{ps}) is 
determined completely by a vector field $S _0$.
However this vector field $S _0$ needs to satisfy (\ref{grule2}).
This kind of vector field $S_0$ is a generalisation of the 
master symmetries defined in \cite{Fuchssteiner1983} 
which here we still call a master symmetry of 
degree $n$ for the more general evolution equation, eqn. (\ref{pee}).
We conclude the discussion above as a theorem. 
\begin{thm} \label{tpds}
Let $\rho$ be a vector field not depending explicitly on the 
time variable $t$. Define 
\be  S_0(\rho) = \rho ,\     
S_{k+1}(\rho)  = \sum_{j=0}^k {k \choose j} [T_j,S_{k-j}(\rho)],
\ k\ge 0,  \label{defs_{k+1}}
\ee
where we assume $T_i=0,\ i\ge m+1$.
If there exists $n\in N$  so that $S_j(\rho)=0,\ j\ge n+1$, then
\be \sigma (\rho) =\sum_{j=0}^n\frac {t^j}{j!}S_j(\rho)\label{defsigmaofrho} 
\ee
is a polynomial-in-time dependent symmetry of the evolution equation,
eqn. (\ref{pee}).
\end{thm}

We shall go on to construct variable-coefficient
integrable equations which possess
higher-degree ptd-symmetries as defined by (\ref{ps}).
We need to start from the centreless Virasoro algebra
\be \left \{ \ba {l} 
\left [K_{l_1},K_{l_2}\right]=0,\ l_1,l_2\ge 0,
\vspace{2mm} \\
\left[K_{l_1},\rho_{l_2}\right]=(l_1+\gamma )K_{l_1+l_2},\ l_1,l_2\ge 0,
\vspace{2mm} \\
\left [\rho_{l_1},\rho_{l_2}\right]=(l_1-l_2)\rho_{l_1+l_2},
\ l_1,l_2\ge 0, \ea
\right. \label{Va} 
\ee
in which 
the vector fields $K_{l_1}=K_{l_1}(u),\ \rho_{l_2}=\rho_{l_2}(u),
\ l_1,l_2\ge 0$, do not 
depend explicitly on the time variable $t$ and $\gamma $ is a fixed constant.
Although the vector fields $\rho_l,\ l\ge 0,$ are not symmetries of
any equations that we want to discuss, an algebra isomorphic to  
this kind of Lie algebra
commonly arises as a symmetry algebra for many well-known continuous and 
discrete integrable equations \cite{ChenL} \cite{MaF} \cite{ChengLB}.
In eqn. (\ref{Va}), the vector fields $\rho_l,\ l\ge 0,$ 
may provide the generators of
Galilean invariance \cite{Fushchych} and 
invariance under scale transformations
for any standard equation $u_t=K_k(u)$.
Let us choose a set of specific vector fields
\be  T_j=K_{i_j}, \ 0\le j\le m,\label{Tchoose}\ee
which yields the following variable-coefficient evolution equation
\be u_t=K_{i_0}+tK_{i_1}+\frac{t^2}{2!}K_{i_2}+\cdots +\frac{t^m}{m!}
K_{i_m}.\label{spee}\ee 
This equation 
still has a hierarchy of time-independent symmetries $K_l, \ l\ge 0,$
and therefore it is integrable in the sense of symmetries \cite{Fokas}. 
What is more, it 
will inherit many integrable properties of $u_t=K_l,\ l\ge0$.
For example, if $u_t=K_l,\ l\ge 0,$ have Hamiltonian structures of the form
\[ u_t=K_l=J\frac {\delta H_l}{\delta u}, \ l\ge 0,\]
where $J$ is a symplectic operator and 
$H_l,\ l\ge 0,$ do not depend explicitly on $t$, 
then the $H_l$
are still conserved densities of eqn. (\ref{spee}) and eqn. (\ref{spee})
is then completely integrable in the commonly used sense for pdes.
In what follows, we need to prove 
that  $\rho_l$ is a master symmetry (as explained above) of 
degree $m+1$ of eqn. (\ref{spee}). 
In fact, according to (\ref{defs_{k+1}}),  
we have 
\[ S_0(\rho_l)=\rho_l,\ 
 S_{k+1}(\rho_l)=[T_k,S_0(\rho_l)]=[K_{i_k},\rho_l]
=(i_k+\gamma )K_{i_k+l},
\ 0\le k\le m,
\]
and further we can prove that $S_j(\rho_l)=0$ when $j\ge m+2$,
which shows that 
$\rho_l$ is a master symmetry of 
degree $m+1$ of eqn. (\ref{spee}). 
Therefore we obtain a hierarchy of ptd-symmetries of the form
\bea  \sigma _l(t,u)&=&\sum_{j=0}^{m+1}\frac{t^j}{j!} S_j(\rho_l)=
\sum_{j=1}^{m+1}\frac{i_{j-1}+\gamma }{j!}t^jK_{i_{j-1}+l}+\rho_l\nonumber\\
&=& \sum_{j=0}^{m}\frac{i_{j}+\gamma }
{(j+1)!}t^{j+1}K_{i_{j}+l}+\rho_l,\ l\ge 0,
\label{sps}
\eea
for the variable-coefficient and integrable equation (\ref{spee}).
Moreover these higher-degree ptd-symmetries 
together with time-independent symmetries $K_l,\ l\ge 0$,
constitute the same centreless Virasoro algebra as
(\ref{Va}), namely
\be \left \{ \ba {l} 
\left [K_{l_1},K_{l_2}\right]=0,\ l_1,l_2\ge 0,
\vspace{2mm} \\
\left[K_{l_1},\sigma_{l_2}\right]=(l_1+\gamma )K_{l_1+l_2},\ l_1,l_2\ge 0,
\vspace{2mm} \\
 \left[\sigma_{l_1},\sigma_{l_2}\right]=(l_1-l_2)\sigma_{l_1+l_2},
 \ l_1,l_2\ge 0.\ea
\right. \label{hVa} 
\ee
For example, we can calculate that
\bea  \left [\sigma_{l_1},\sigma_{l_2}\right] & = &
\Bigl[\sum_{j=0}^m\frac{i_j+\gamma }{(j+1)!}t^{j+1}K_{i_j+l_1}+\rho_{l_1},
\sum_{j=0}^m\frac{i_j+\gamma }{(j+1)!}t^{j+1}K_{i_j+l_2}+\rho_{l_2}\Bigr]
\nonumber\\
&=&    
\Bigl[\sum_{j=0}^m\frac{i_j+\gamma }
{(j+1)!}t^{j+1}K_{i_j+l_1},\rho_{l_2}\Bigr]
+ \Bigl[\rho_{l_1},
\sum_{j=0}^m\frac{i_j+\gamma }{(j+1)!}t^{j+1}K_{i_j+l_2}\Bigr]
+\left[\rho_{l_1},\rho_{l_2}\right]
\nonumber \\
&=& \sum_{j=0}^m\frac{(l_1-l_2)(i_j+\gamma )}{(j+1)!}t^{j+1}K_{i_j+l_1+l_2}
+(l_1-l_2)\rho_{l_1+l_2}\nonumber\\
&& = (l_1-l_2)\sigma_{l_1+l_2}. \nonumber
\eea
The algebra (\ref{hVa}) also gives us 
a new realisation of centreless Virasoro algebras.
By now we may very much see that there exist 
higher-degree ptd-symmetries for some 
evolution equations in $1+1$ dimensions. Moreover our derivation
does not refer to any particular choices of dimensions and space variables.
Hence the evolution equation (\ref{spee}) may be not only both
continuous and discrete, but also both $1+1$ and $2+1$ dimensional.

Actually there are many 
integrable equations which possess a centreless Virasoro 
algebra (\ref{Va}) (see \cite{ChenL} \cite{MaF} \cite{ChengLB} 
\cite{Ma1990} \cite{OevelFZ}
for example). 
Among the most famous examples are the KdV hierarchy
in the continuous case and the Toda lattice hierarchy
in the discrete case.
Through the theory above, we can say that
a KdV-type equation 
\be u_t=tK_0+K_1=tu_x+u_{xxx}+6uu_x\ee
possesses a hierarchy of second-degree time-polynomial-dependent symmetries
\be 
\sigma _l=\frac 32 tK_{l+1}+\frac 14t^2K_{l}+\rho_l,\ l\ge 0,
\ee
where the vector fields $K_l,\,\sigma _l,\ l\ge0,$ are defined by
\[ K_l=\Phi^l u_x,\ \rho_{l}=\Phi ^l(u+\frac 12 xu_x), \ \Phi = \part ^2+
4u+2u_x\part ^{-1},\ l\ge 0.\]
They constitute a centreless Virasoro algebra (\ref{Va}) with $\gamma =
\frac 12 $ \cite{Ma1990} \cite{Ma1992} and thus so do the symmetries 
$K_l,\, \sigma_l,\ l\ge 0$. 
We can also conclude that a Toda-type lattice equation 
\bea && (u(n))_t=\left(\ba {c} p(n) \vspace{2mm} \\ v(n)   \ea  \right) _t
=K_0+tK_1+\frac{t^2}{2!}K_0\nonumber\\
&& =(1+\frac 12 t^2)\left(\ba {c} v(n)-v(n-1) \vspace{2mm}\\     
v(n)(p(n)-p(n-1))\ea  \right)\nonumber \\
&& \quad 
+ t\left(\ba {c} p(n)(v(n)-v(n-1))+v(n)(p(n+1)-p(n-1)) \vspace{2mm}\\     
v(n)(v(n-1)-v(n+1))+v(n)(p(n)^2-p(n-1)^2)\ea  \right)
\eea
possesses a hierarchy of third-degree time-polynomial-dependent symmetries
\be 
\sigma _l=tK_l+t^2K_{l+1}+\frac 16 t^3 K_l+\rho _l, \ l\ge 0,\ee
where the corresponding vector fields are defined by
\bea &&
K_l=\Phi ^{l}K_0,\ K_0=\left ( \ba {c} v-v^{(1)}\vspace{1mm}
\\ v(p-p^{(-1)})\ea \right ),
\ l\ge0 , \nonumber
\\ && \rho_{l}=\Phi ^l \rho_0,
\  \rho_0=\left ( \ba {c} p\vspace{2mm}
\\ 2v\ea \right ),  \ l\ge 0,   \nonumber
 \eea 
 in which  the hereditary operator  $\Phi$ is defined by
\[ \Phi=\left ( \ba {cc} p&(v^{(1)}E^2-v)(E-1)^{-1}v^{-1}\vspace{1mm}
\\  v(E^{-1}+1)& v(pE-p^{(-1)})(E-1)^{-1}v^{-1}\ea \right ).
\]
Here we have used a normal shift operator $E$: $(Eu)(n)=u(n+1)$ and 
$u^{(m)}=E^mu$.
These discrete 
vector fields $K_l,\ l\ge 0$, (see \cite{Tu} for more information)
together with the discrete vector fields $\rho _l,\ l\ge 0$, 
constitute a centreless Virasoro algebra (\ref{Va})
with $\gamma =1$ \cite{MaF} and the symmetry 
Lie algebra of $\sigma _l,\
l\ge 0$ and $K_l,\ l\ge 0$, has the same commutation relations as that
Virasoro algebra.

More generally, we can consider further algebraic structures by starting 
from a more general graded Lie algebra.
In keeping with the notation in \cite{Ma1992a},
let us write a graded Lie algebra consisting 
of vector fields not depending explicitly
on the time variable $t$ as follows:
\be E(R)=\sum_{i=0}^\infty E(R_i),\ [E(R_i), E(R_j)]\subseteq E(R_{i+j-1}),
\ i,j\ge 0,\label{gradeda}\ee 
where $E(R_{-1})=0$. Note that such a graded Lie algebra is called a master
Lie algebra in \cite{Ma1992a} since it is actually not 
a graded Lie algebra as defined in \cite{Kac}. 
However we still call it a graded Lie algebra because it is very similar.
Choose 
\be T_i=K_i\in E(R_0) ,\ 0\le i\le m,\label{Tchoose2}\ee
 and consider a 
variable-coefficient evolution equation
\be u_t=\sum_{i=0}^m \frac {t^i}{i!}T_i=K_0+tK_1+\frac {t^2}{2!}K_2+\cdots
+\frac{t^m}{m!}K_m.\label{spee2}\ee 
Before we state the main result, we derive two properties of 
the generating vector fields $S_j,\ j\ge 0.$

\begin{lemma}
Assume that $T_i,\ 0\le i\le m,$ are defined by (\ref{Tchoose2}),
and let $l\ge 0$ and $\rho_l\in E(R_l)$. Then the vector fields $S_j(\rho_l),
\ j\ge 0$, defined by (\ref{defs_{k+1}}) satisfy the following property
\bea && S_{(\al -1)(m+1)+\beta }(\rho_l)\in \sum_{i=0}^{l-\al }E(R_i),
\ 1\le \al \le l,\ 1\le \beta \le m+1, \label{sigmaprop1}\\
&& S_j(\rho_l)=0,\ j\ge l(m+1)+1.  \label{sigmaprop2}
\eea  \label{sigmaproplemma}
\end{lemma}
{\bf Proof:} 
Note the definition (\ref{defs_{k+1}}) of $S_{j}(\rho_l),\ j\ge 0$, and 
$T_i=K_i,\ 0\le i\le m$.
We can calculate that
\bea && S_{\al (m+1)+\beta +1}(\rho_l)=\sum_{\gamma=0 }^m
{ \al (m+1)+\beta \choose \gamma}
\left [K_{\gamma },S_{\al (m+1)+\beta 
-\gamma}(\rho_l) \right ]\nonumber \\
&& =\sum_{\gamma =0}^{\beta -1} 
{\al (m+1)+\beta \choose \gamma }
\left [K_{\gamma },S_{\al (m+1) 
+\beta -\gamma}(\rho_l) \right ]\nonumber \\
&&\quad +\sum_{\gamma =\beta }^m 
{ \al (m+1)+\beta \choose \gamma} 
\left [K_{\gamma },S_{(\al -1) (m+1) 
+[(m+1)-(\gamma-\beta )]}(\rho_l) \right ]\nonumber \\
&& \in \sum_{i=0}^{l-(\al +2)} E(R_i)+\sum_{i=0}^{l-(\al +1)}
E(R_i)=\sum_{i=0}^{l-(\al +1)} E(R_i),\nonumber
\eea
where in the last but one step we have used the induction assumption. 
This result shows that (\ref{sigmaprop1}) is true by mathematical induction. 
The proof of 
(\ref{sigmaprop2}) is the same so that the proof of the Lemma is complete.
$\vrule height 3mm width 1mm depth 0.4mm$

\begin{lemma} \label{lielemma} 
Assume that $T_i,\ 0\le i\le m,$ are defined by (\ref{Tchoose2}),
and let $l_1,l_2\ge 0$ and $\rho_{l_1}\in E(R_{l_1}),\ 
\rho_{l_2}\in E(R_{l_2})$. Then we have 
\be S_k([\rho_{l_1},\rho_{l_2}])=\sum_{i+j=k}{k\choose i}
[S_i(\rho_{l_1}),S_j(\rho_{l_2})],\ k\ge 0,\label{s_kprodp}\ee
where the $S_j(\rho),\ j\ge 0$, are defined by (\ref{defs_{k+1}}).
\end{lemma}
{\bf Proof:} 
We use mathematical induction to prove the required result.
Noting that $T_i=K_i,\ 0\le i\le m,$ we can calculate that
\bea && S_{k+1}([\rho_{l_1},\rho_{l_2}])= \sum_{i+j=k}{k\choose i}\Bigl[
K_i,S_j([\rho_{l_1},\rho_{l_2}])\Bigr] \nonumber\\
&&= \sum_{i+j=k}{k\choose i}\Bigl[
K_i,\sum_{\al +\beta =j}{j \choose \al }\left[S_\al (\rho_{l_1}),
S_\beta (\rho_{l_2})\right]\Bigr]
\ \quad {\rm (by\ the \  induction\  assumption)}\nonumber\\
&&= \sum_{i+j=k}{k\choose i}\sum_{\al +\beta =j}{j \choose \al }
\Bigl[K_i, \left[S_\al (\rho_{l_1}),
S_\beta (\rho_{l_2})\right]\Bigr] \nonumber\\
&&= \sum_{i+\al +\beta =k}\frac {k!}{i!\al !\beta !}\Bigl[
K_i,\left[S_\al (\rho_{l_1}),
S_\beta (\rho_{l_2})\right]\Bigr] \nonumber\\
&&= \sum_{i+\al +\beta =k}\frac {k!}{i!\al !\beta !}\left \{\Bigl[
\left[K_i,S_\al (\rho_{l_1})\right],
S_\beta (\rho_{l_2})\Bigr]+ \Bigl [ S_\al (\rho_{l_1}),\left [ K_i, S_\beta 
(\rho_{l_2})\right] \Bigr ]\right \}\nonumber \\
&&= \sum_{j +\beta =k}{k\choose j} \Bigl[
\sum_{i+\al =j}{j\choose i}\left [K_i,S_\al (\rho_{l_1})\right],
S_\beta (\rho_{l_2})\Bigr] \nonumber\\
&& \qquad +\sum_{\al +j=k}{k \choose j }\Bigl [S_\al (\rho_{l_1}),
\sum_{i+\beta =j}{j\choose i}\left[K_i,
S_\beta (\rho_{l_2})\right]\Bigr] \nonumber\\
&& =\sum_{j+\beta =k}{k\choose j} \left [S_{j+1}(\rho_{l_1}),S_{\beta}
(\rho_{l_2})\right]+
\sum_{\al +j =k}{k \choose j } \left [S_{\al }(\rho_{l_1}),S_{j+1}
(\rho_{l_2})\right]\nonumber \\
&& =\sum_{i+j=k+1}{k+1\choose i} \left[ S_i(\rho_{l_1}), S_j(\rho_{l_2})
\right],\ k\ge 0,\nonumber
\eea
and this yields the key step in the mathematical induction. On the other 
hand, we easily see that
\[ S_0([\rho_{l_1},\rho_{l_2}])=[\rho_{l_1},\rho_{l_2}]
=[S_0(\rho_{l_1}),S_0(\rho_{l_2})].\]
Therefore mathematical induction gives the proof of the equality
(\ref{s_kprodp}). 
$\vrule height 3mm width 1mm depth 0.4mm$

\begin{thm}
Assume that $T_i,\ 0\le i\le m,$ are defined by (\ref{Tchoose2}),
and let $l\ge 0$ and $\rho_l\in E(R_l)$. Then the vector field
\be \sigma (\rho_l)=\sum_{j=0}^{l(m+1)}\frac{t^j}{j!}S_j(\rho_l),
\label{sigmarhol}\ee
where the 
$S_j(\rho_l), \ 0\le j\le l(m+1)$, are defined by (\ref{defs_{k+1}}),
is a time-independent symmetry of (\ref{spee2}) when $l=0$ and 
a polynomial-in-time dependent symmetry of (\ref{spee2}) when $l>0$. 
Furthermore we have
\be [\sigma (\rho_{l_1}), 
\sigma (\rho_{l_2})]=\sigma ([\rho_{l_1},\rho_{l_1}]),\ \rho_{l_1}\in 
E(R_{l_1}),\ \rho_{l_2}\in E(R_{l_2}), \ l_1,l_2\ge 0, \label{liep}
\ee
and thus all symmetries $\sigma (\rho_l)$ with $\rho_l\in E(R_l),\ l\ge 0$,
constitute the same graded Lie algebra as (\ref{gradeda}) and
the map $\sigma: \rho_l\mapsto \sigma(\rho_l)$ is a Lie homomorphism
between two graded Lie algebras $E(R)$ and $\sigma(E(R))$.
\end{thm}
{\bf Proof:} By Lemma \ref{sigmaproplemma}, we can observe that
$\sigma (\rho_l)$ defined by (\ref{sigmarhol}) is a symmetry of 
(\ref{spee2}). We go on to prove (\ref{liep}).  Assume that 
$\rho_{l_1}\in E(R_{l_1}),\  \rho_{l_2}\in E(R_{l_2}),\ l_1,l_2\ge 0$.
By Lemma \ref{sigmaproplemma} 
and Lemma \ref{lielemma}, we can make the following
calculation
\bea && \left [\sigma (\rho_{l_1}),\sigma (\rho_{l_2}) \right]
=\Bigl [\sum_{i=0}^{l_1(m+1)}\frac {t^i}{i!}S_i(\rho_{l_1}), 
\sum_{j=0}^{l_2(m+1)}\frac {t^j}{j!}S_j(\rho_{l_2})\Bigr]
\nonumber \\
&& =\sum_{k=0}^{(l_1+l_2-1)(m+1)}\frac {t^k}{k!}\sum_{i+j=k}
{k \choose i}\left[ S_i(\rho_{l_1}), S_j(\rho_{l_2})\right] \ \ 
\ {\rm (by\  Lemma\  \ref{sigmaproplemma})}\nonumber\\
&& =\sum_{k=0}^{(l_1+l_2-1)(m+1)}\frac{t^k}{k!} S_k([\rho_{l_1},\rho_{l_2}])
\ \ \ {\rm (by\  Lemma\  \ref{lielemma})}\nonumber \\
&& = \sigma ([\rho_{l_1},\rho_{l_2}]).\nonumber
\eea
The rest is then obvious and the required result is obtained.
$\vrule width 1mm height 3mm depth 0.4mm$ 

A graded Lie algebra has been exhibited for the time-independent 
KP hierarchy \cite{OevelF}
in \cite{Fuchssteiner1983}
\cite{Ma1992a}, and it includes a centreless Virasoro algebra 
\cite{ChengLB} \cite{ChenLL}. 
The ordinary time-independent KP equation being considered here 
is the following
\be u_t=\part ^{-1}_xu_{yy}-u_{xxx}-6uu_x .\label{KP1}\ee 
From this we may now go on to generate the corresponding 
graded Lie algebra of ptd-symmetries
for a resulting new set of variable-coefficient KP equations, but in this 
connection the reader must be referred to the comparable analysis 
in \cite{MaBC} mentioned below.

The idea of using graded Lie algebras as described in this paper 
is rather similar to the thinking
used to extend the inverse scattering transform 
from $1+1$ to higher dimensions
\cite{Caudrey}. Moreover the resulting symmetry algebra 
consisting of  the $\sigma(\rho_l), l\ge 0$, provides a new realisation 
of a graded Lie algebra (\ref{gradeda}). The theory
also shows us that more information can be extracted from  
graded Lie algebras, which is itself very interesting. What is more,   
we have shown here that there do exist various integrable equations 
in $1+1$ dimensions, such as KdV-type equations,
possessing higher-degree polynomial-in-time dependent symmetries.
We report a graded Lie algebra of ptd symmetries for a corresponding 
new set of variable coefficient {\it modified} KP equations in a second 
article \cite{MaBC}. In \cite{MaBC} we display this modified KP 
hierarchy explicitly, the time independent modified KP equation being, 
in comparison with (\ref{KP1}), the equation
\be u_t=\frac14 u_{xxx}-\frac 38 u^2u_x-\frac 34 u_x\part ^{-1}_xu_y+\frac 34
\part ^{-1}_x u_{yy} .\label{KP2}\ee
In \cite{MaBC} we show also that this hierarchy actually has {\it two} 
Virasoro algebras and {\it two} graded Lie algebras.

We also hope to show elsewhere the connections between the rather
general algebraic 
structure established in this paper and the specific representation
of the $W_\infty $ and $W_{1+\infty}$ algebras developed in connection
with two-dimensional quantum gravity as described in in Refs. 
\cite{BulloughC1} \cite{BulloughC2}
(In \cite{BulloughC1} \cite{BulloughC2}, 
these two infinite dimensional algebras
were developed for the ordinary KP hierarchy and included the algebra,
containing the centreless Virasoro algebra, of Ref.\cite{ChengLB}). 
In this connection, we note already that if, for example,
$E(R_i)={\rm span}\{A_{im}|\,  m\ge 1\},\ i\ge 0,$ and we impose
\[ [A_{im},A_{jn}]=\sum_{l={\rm min}(i-1,j-1)}^{i+j-2}a_l(i-1,j-1,
m-1,n-1)A_{l+1,m+n-1},\]
where the coefficients $a_l$ are defined by
\[ \Bigl [ x^{i+m+1}\frac {d^{i+1}}{dx^{i+1}}, 
x^{j+n+1}\frac {d^{j+1}}{dx^{j+1}}  \Bigl]= 
\sum_{l={\rm min}(i,j)}^{i+j}a_l(i,j,m,n)x^{l+m+n+1}\frac {d^{l+1}}{dx^{l+1}}
,\]
then the $E(R)=\sum_{i=0}^\infty E(R_i)$ 
is a sub-algebra of the $W_{1+\infty}$ algebra of 
Refs. \cite{BulloughC1} \cite{BulloughC2} by the identification 
$A_{im}=\tau_{i-1,m-1}$; here the $\tau_{i-1,m-1}$ are the elements forming the 
$W_{1+\infty}$ algebra \cite{BulloughC1} \cite{BulloughC2} and they may be 
realized by $x^{i+m-1}\frac{d^i}{dx^i}$. 

\noindent{\bf Acknowledgements:} 
One of the authors (WXM) 
would like very much to thank the 
Alexander von Humboldt Foundation for 
the financial support, which 
made his visit to UMIST possible.
He is also grateful to Prof. B. Fuchssteiner 
for his kind and stimulating discussions.

\end{document}